\documentclass[10pt,twocolumn,superscriptaddress,prl,showpacs,floatfix]{revtex4}
\usepackage{dcolumn}                 % Aligning table columns
\usepackage{times}
\usepackage{nicefrac}
\usepackage{amsmath}
\usepackage{amsfonts}
\usepackage{amssymb}
\usepackage{amsthm}
\usepackage{xcolor}
\usepackage{graphicx}

\newcommand*{\cent}[1]{\multicolumn{1}{c}{$#1$}}
\newcolumntype{w}[1]{D{.}{.}{#1}}

\newcommand{\icm}{\mathrm{cm}^{-1}}

\definecolor{green}{rgb}{.0,.5,.0}

\definecolor{red}{rgb}{1.,.0,.0}

\definecolor{grey}{rgb}{0.5,0.5,0.5}

\begin{document}
\preprint{Version 1.1}

\title{Explicitly correlated wave function for a boron atom} 

\author{Mariusz Puchalski}
\affiliation{Faculty of Chemistry, Adam Mickiewicz University,
             Umultowska 89b, 61-614 Pozna{\'n}, Poland}

\author{Jacek Komasa}
\affiliation{Faculty of Chemistry, Adam Mickiewicz University,
             Umultowska 89b, 61-614 Pozna{\'n}, Poland}

\author{Krzysztof Pachucki}
\affiliation{Faculty of Physics, University of Warsaw, Pasteura 5, 02-093 Warsaw, Poland}

\date{\today}

\begin{abstract}
We present results of high-precision calculations for a boron atom's properties 
using wave functions expanded in the explicitly correlated Gaussian basis.
We demonstrate that the well-optimized 8192 basis functions enable a determination 
of energy levels, ionization potential, and fine and hyperfine splittings in atomic 
transitions with nearly parts per million precision. 
The results open a window to a spectroscopic determination of nuclear properties 
of boron including the charge radius of the proton halo in the $^8$B nucleus.   
\end{abstract}

\maketitle

\section{Introduction}
While for hydrogenic ions the nonrelativistic wave function is known
exactly, for all larger atomic systems it has to be obtained numerically, 
most often with the help of the variational principle.
The numerical precision achieved for a few electron systems can, nevertheless, 
be very high. For example, nonrelativistic energies of the He atom are known with 
more than 20 digits of accuracy \cite{PhysRevA.65.054501, PhysRevA.66.024501,Schwartz:06}, 
of Li with 15 digits \cite{Wang:12,Puchalski:09}, and very recently
the precision achieved for the Be atom reached 11 significant digits 
\cite{PhysRevA.80.022514,Puchalski:13a,Puchalski:14a}.
The computational approach employed in all those atomic studies is based on explicitly correlated functions 
of the exponential or Gaussian form, which are the best known representations 
of the nonrelativistic wave function. For three-electron systems, the most accurate solution of 
the Schr\"odinger equation is obtained with the Hylleraas (exponential times polynomial)
basis functions \cite{Puchalski:06a,Puchalski:09,Wang:12}. In such systems, the accuracy 
of the theoretical predictions for transition energies and isotope shifts is limited
by the approximate treatment of higher-order $\sim\!m \alpha^{6,7}$ QED
corrections rather than by numerical inaccuracies of the nonrelativistic wave function.  
Methods with Hylleraas functions have been extended to four-electron atomic systems 
but only for some restricted selection of basis functions,
because of significant difficulties in evaluation of matrix elements \cite{Busse:98,King:11}. 
Even more difficult integrals appear in the matrix elements of relativistic operators.
Unquestionably, significant efforts have to be made to improve the Hylleraas approach, 
in order for it to be practical for the four and more electron systems.
Therefore, at present, the method of choice for such systems is that based on
explicitly correlated Gaussian (ECG) functions.
The effectiveness of the ECG functions in treating few-electron problems has already 
been demonstrated by high-precision calculations of the nonrelativistic energies of atomic 
and molecular systems \cite{Komasa:02a,Bachorz:09,Komasa:11,Przybytek:12,Bubin:13,Mitroy:13}. 
In particular, for the beryllium atom 
the highest accuracy has been obtained using the ECG functions 
\cite{Komasa:95,Komasa:01b,Stanke:07,Bubin:12,Puchalski:13a}. 
The main advantage of the ECG method is that the underlying integration is manageable 
and very fast in numerical evaluation due to the compact formulas involving only
elementary functions. 
On the other hand, the Gaussian functions have the drawback of improper asymptotic behavior
since they decay too fast at long inter-particle distances. 
They also have an incorrect short-range form and fail to correctly describe the Kato cusp. 
However, these two flaws can be overcome if one employs a sufficiently large and
well-optimized ECG basis set. The issue is subtler in calculations 
of relativistic and QED properties, in which the local inaccuracies of the wave functions 
result in significant numerical loss of mean values. 
One has to carefully optimize over a large number (oftentimes exceeding $10^5$) of variational 
parameters matching local behavior of the exact wave function
and employ dedicated techniques which accelerate the convergence of nearly singular matrix elements
\cite{PCK05}. 
An additional drawback of the approach based on fully correlated functions
is the cost resulting from antisymmetrization of the wave function
which grows like $N!$ with the number of electrons $N$. 

In this paper we demonstrate
that in spite of this high evaluation costs, the methods based on ECG functions may give
a spectroscopic accuracy for five-electron systems, such as the boron atom, in a realistic 
computational time. We report on the calculation of nonrelativistic energies of the ground $2^2P$ and
the excited $3^2S$ levels, and the leading relativistic corrections including 
the fine and hyperfine structure of the ground state. The achieved numerical accuracy
is several orders higher than those of any previous calculations and not always in agreement with them.
In addition, we provide accurate results for a four-electron B$^+$ ion needed to determine 
the ionization potential of the boron ground state. 

\section{Nonrelativistic Hamiltonian and corrections}

The determination of accurate wave functions corresponding to 
the nonrelativistic, clamped nucleus Hamiltonian (in natural units)
%
%\begin{equation}
%H_0 = \sum_a\frac{p_a^2}{2 m} + \frac{p_N^2}{2 m_N}-\sum_a\frac{Z\,\alpha}{r_a}
%+\sum_{a>b} \frac{\alpha}{r_{ab}}. \label{h0}
%\end{equation} 
\begin{equation}
H_0 = \sum_a\frac{p_a^2}{2\,m}-\sum_a\frac{Z\,\alpha}{r_a}
+\sum_{a>b} \frac{\alpha}{r_{ab}}. \label{h0}
\end{equation} 
is the main subject of this paper. If the wave function $\Phi$ is determined,  
all the corrections to the energy $E_0$
in the following perturbative expansion in the fine-structure constant $\alpha \sim 1/137$
\begin{equation}
E = E^{(2)}+E^{(4)} + E^{(5)} + E^{(6)} + \ldots,\,\quad E^{(n)}\sim m \alpha^n
\label{eq:expansion}
\end{equation}
can be expressed in terms of expectation values 
$\langle\Phi|\ldots|\Phi\rangle \equiv \langle \ldots \rangle$ of known operators.
Complete nonrelativistic energy $E^{(2)}$ consists of the clamped nucleus energy $E_0$ 
and the kinetic energy of the nucleus $H_\mathrm{N}$, which can be calculated 
in the center of mass frame as a small perturbation, 
\begin{equation}
H_\mathrm{N} =\frac{p_\mathrm{N}^2}{2 \,m_\mathrm{N}} = \frac{m}{m_\mathrm{N}} \bigg( \sum_a \frac{p_a^2}{2\,m} +   \sum_{a<b} \frac{\vec p_a \cdot \vec p_b}{m} \biggr)\,,
\end{equation}
where $m_\mathrm{N}$ is the nuclear mass. The leading relativistic $E^{(4)}$ correction 
is calculated as the mean value of the Breit-Pauli Hamiltonian given by
\begin{eqnarray}
H^{(4)} &=& H_{\rm ns} + H_{\rm fs} + H_{\rm hfs}\,,\label{H4}\\
H_{\rm ns} &\equiv& \sum_a\,\left[ -\frac{p_a^4}{8\,m^3}+
\frac{Z\,\alpha\,\pi}{2\,m^2}\,\delta^3(r_a)\right] \\ 
&& \hspace{-0.5cm} +\sum_{a > b}\left[
\frac{\pi\,\alpha}{m^2}\,\delta^3(r_{ab})
-\frac{\alpha}{2\,m^2}\,p_a^i\biggl(
\frac{\delta^{ij}}{r_{ab}}+\frac{r^i_{ab}\,r^j_{ab}}{r^3_{ab}}\biggr)\,p_b^j\right], \nonumber \\
H_{\rm fs} &\equiv& \sum_a\,\frac{Z\,\alpha}{4\,m^2\,r_a^3}\,\Big[(g-1) 
\vec{r}_a\times\vec{p}_a\Big]\cdot\vec{\sigma}_a  \nonumber \\
&& \hspace{-0.5cm} +\sum_{a\neq b}\, \frac{\alpha}{4\,m^2\,r_{ab}^3}\,
\vec{\sigma}_a \cdot \Big[g\,\vec{r}_{ab} \times\vec{p}_b-(g-1)\,\vec{r}_{ab} \times\vec p_a\Big]
\nonumber \\
&& + \sum_{a>b} \frac{\alpha\,g^2}{16\,m^2\,r_{ab}^3}\,\sigma_a^i\,\sigma_b^j\,\biggl(
\delta^{ij}-\frac{3\,r_{ab}^i\,r_{ab}^j}{r_{ab}^2}\biggr). 
\end{eqnarray}
The leading order Hamiltonian for the hyperfine splitting  is
\begin{eqnarray}
H_{\rm hfs} &\equiv&
\sum_a\biggl[
\frac{1}{3}\,\frac{Z\,\alpha\,g\,g_{\rm N}}{m\,m_{\rm N}}\,
\vec \sigma_a\cdot\vec I\,\pi\,\delta^3(r_a)
\nonumber \\ &&
+\frac{Z\,\alpha\,g_N}{2\,m\,m_N}\,\vec I\cdot\frac{\vec r_a}{r_a^3}\times\vec p_a
\nonumber \\ &&
-\frac{Z\,\alpha\,g\,g_N}{8\,m\,m_N}\,
\frac{\sigma_a^i\,I^j}{r_a^3}\,
\biggl(\delta^{ij}-3\frac{r_a^i\,r_a^j}{r_a^2}\biggr)
\nonumber \\ &&\hspace{-1ex}
%-\frac{Z\,\alpha\,(g_N-1)}
%{2\,m_N^2}\,\vec I\cdot\frac{\vec r_a}{r_a^3}\times\vec p_N
%\nonumber \\ &&
+\frac{Q_\mathrm{N}}{6}\,\frac{\alpha}{r_a^3}\,
\biggl(\delta^{ij}-3\frac{r_a^i\,r_a^j}{r_a^2}\biggr)\,
\frac{3\,I^i\,I^j}{I\,(2\,I-1)}\biggr]\,, 
\end{eqnarray}
where $\vec{I}$ is the nuclear spin, $g$ is the electron $g$ factor, 
$Q_\mathrm{N}$ is the electric quadrupole moment of the nucleus, 
and $g_{\rm N}$ is the nuclear $g$ factor 
\begin{equation}
g_{\rm N} = \frac{m_{\rm N}}{Z\,m_{\rm p}}\,\frac{\mu}{\mu_{\rm N}}\,\frac{1}{I}.
\end{equation}
It is convenient to rewrite the expectation value of the hyperfine splitting $H_{\rm hfs}$ Hamiltonian in 
terms of commonly used $A_J$ and $B_J$ coefficients,
\begin{eqnarray}
\langle H_{\rm hfs} \rangle_J &=&
A_J\,\vec I\cdot\vec J + \frac{B_J}{6}\,
\frac{3\,(I^i\,I^j)^{(2)}}{I\,(2\,I-1)}\,
\frac{3\,(J^i\,J^j)^{(2)}}{J\,(2\,J-1)},
\end{eqnarray}
where $\vec J$ is the total electronic angular momentum.
For this purpose we decompose $H_{\rm hfs}$
\begin{equation}
H_{\rm hfs} = 
\vec I\cdot \vec G + \frac{G^{ij}}{6}\,\frac{3\,(I^i\,I^j)^{(2)}}{I\,(2\,I-1)},
\end{equation}
and these coefficients are
\begin{eqnarray}
A_J &=& \frac{1}{J\,(J+1)}\,\langle \vec J\cdot\vec G\rangle\,,\\
B_J &=& \frac{2}{(2\,J+3)\,(J+1)}\,\langle J^i J^j\,G^{ij}\rangle_J\,.
\end{eqnarray}

Higher-order corrections to the atomic energy originate from QED. 
They are significantly smaller then $E^{(4)}$
because of the higher powers of $\alpha$:
\begin{eqnarray}\label{eq:HQED}
E^{(5)} &=& \frac{4\,Z\,\alpha^2}{3\,m^2}\,\left[\frac{19}{30}+\ln(\alpha^{-2}) - \ln k_0\right]\,
\sum_a\,\langle \delta^3(r_a) \rangle  \nonumber \\ 
&& + \frac{\alpha^2}{m^2} \left[\frac{164}{15}+\frac{14}{3}\,\ln\alpha
\right]\,\sum_{a<b} \,\langle \delta^3(r_{ab}) \rangle  \\
&& -\frac{7}{6\,\pi} \,m\,\alpha^5 \sum_{a<b} \,\biggl \langle P\left(\frac{1}{m\,\alpha\,r_{ab}^3} \right) \biggr \rangle\,,  \nonumber\\
E^{(6)} &\approx& \frac{\pi\,Z^2\,\alpha^3}{m^2}\,\left[\frac{427}{96} - 2 \ln(2) \right] \sum_a \langle \delta^3(r_a) \rangle.
\end{eqnarray}

\section{Reduction of matrix elements}

We represent the wave function $\Phi^i$ of the five-electron $^2P$ atomic state in the form 
\begin{equation}
\Phi^i = \frac{1}{\sqrt{5!}}\,{\cal A} \bigg[\sum_n t_n \phi_n^i(\{ \vec r_a\})\,\chi_{\{a\}}\bigg]
\end{equation}
where $t_n$ is a linear coefficient, $\chi_{\{a\}}$ is the spin wave function
\begin{equation}
\chi_{\{a\}} = (\alpha_1\,\beta_2-\beta_1\,\alpha_2)\,(\alpha_3\,\beta_4-\beta_4\,\alpha_3)\,\alpha_5
\end{equation}
and $\{a\}$ and $\{\vec r_a\}$ denote the sequences $1,2,3,4,5$ 
and $\vec r_1, \vec r_2, \vec r_3, \vec r_4, \vec r_5$, respectively. The symbol
$\cal A$ denotes antisymmetrization and $\phi^i(\{ \vec r_a\})$ is a spatial function with Cartesian
index $i$ that comes from one of the electron coordinates
\begin{equation}
\phi^i(\{ \vec r_k\})=r_m^i\exp{\left[-\sum_{k=1}^N a_k\,r_k^2-\sum_{l>k=1}^N b_{kl}\,r_{kl}^2\right]}.
\end{equation}

The normalization we assume is
\begin{equation}
\sum_i \langle\Phi^i|\Phi^i \rangle = \sum_n\sum_m t_n^* t_m\sum_i
\bigl\langle \phi_n^i | \phi_m^i \bigr\rangle_S = 1 \label{norm}
\end{equation}
where
\begin{equation}
\bigl\langle \phi'^{\,i} | \phi^i \bigr\rangle_S = 
\bigl\langle \phi'^{\,i}(\{\vec r_b\})|{\cal P}_{\{a\}}
             [c_{\{a\}}\,\phi^i(\{ \vec r_a\})] \bigr\rangle
\end{equation}	
and where ${\cal P}_{\{ a \}}$ denotes the sum over all permutations of $\{a\}$.
From now on we will assume that a repeated Cartesian index is implicitly summed up.
We introduce also another type of a matrix element, which will be used later
\begin{equation}
\bigl\langle\phi'^{\,i} | Q | \phi^k \big \rangle_F =   
\sum_c\bigl\langle\phi'^{\,i}(\{\vec r_a\})|
Q_c\,{\cal P}_{\{b\}} \big[c^{Fc}_{\{b\}}\,\phi^k(\{\vec r_b\})\big] \big \rangle\,.
\end{equation}
The $c$ coefficients are integers and depend on the permutation $\{a\}$. 
Since the number of permutations is $5! = 120$, we cannot explicitly write them down here. 
What is important is that all the matrix elements are either of the standard form $\langle\ldots\rangle_S$
with the constant coefficients $c$ 
or of the Fermi interaction form $\langle\ldots\rangle_F$ with the $c^F$ coefficients. 

The $^2P_{1/2}$ and $^2P_{3/2}$ wave functions are constructed using
Clebsch-Gordan coefficients. Expectation values with these wave functions
can be reduced to spinless expressions with an algebraic prefactor $K_J$ 
for $J=1/2$ and $3/2$. Namely, for an operator $Q$, the first-order matrix elements 
with an auxiliary notation $\{K_{1/2},K_{3/2}\}$ take the form 
\begin{equation}\label{E:Q}
\langle\Phi|Q|\Phi \rangle =  \{1,1\} \, \bigl\langle \phi^i|   
Q |\phi^i \bigr\rangle_S\,,  \\
\end{equation}
\begin{equation}
 \langle \Phi |\sum_c \vec \sigma_c \cdot \vec Q_c |\Phi \rangle =   
\{1,-1/2\} \, \imath \,\epsilon^{ijk}  
\bigl\langle\phi^i | Q^j | \phi^k \big \rangle_F\,,
\end{equation}
\begin{equation}
\frac{1}{J(J+1)}\langle \Phi |\vec J  \cdot \sum_c \vec \sigma_c \, Q_c |\Phi \rangle
= \{-2/3,2/3\} \bigl\langle\phi^i | Q | \phi^i \big \rangle_F\,,
\end{equation}
\begin{equation}
\frac{1}{J (J+1)}\, \langle \Phi |\vec J  \cdot \vec Q |\Phi \rangle = 
\{-2/3,-1/3\} \,\imath\,\epsilon^{ijk} \bigl\langle\phi^i |
Q^j | \phi^k \big \rangle_S\,,
\end{equation}
\begin{eqnarray}
\frac{1}{J (J+1)} \langle \Phi|J^i \sum_c \sigma_c^j \,Q_c^{ij} |\Phi \rangle &=& 
\{4/3,-2/15\} \nonumber \\ 
&& \hspace{-2ex} \times \bigl\langle\phi^i | Q^{ij} | \phi^j  \big \rangle_F\,,
\end{eqnarray}
\begin{eqnarray}\label{E:Qij}
\frac{2}{(2\,J+3)\,(J+1)}\,\langle \Phi |J^i\,J^j\,Q^{ij} |\Phi \rangle &=& 
\{0,-1/5\} \nonumber \\ 
&& \hspace{-6ex} \times \bigl\langle\phi^i | Q^{ij} | \phi^j \big \rangle_S\,.
\end{eqnarray}
The above spin reduced matrix elements involve only scalars built of spatial variables $\vec r_a$,
and therefore they all can easily be expressed by Gaussian type integrals.

\section{Numerical calculations and results}

In the numerical calculations we employed the ECG basis functions of progressively doubled size
from $1024$ to $8192$ terms for the B atom, and from $512$ to $4096$ terms for the B$^+$ ion. 
The nonlinear parameters were optimized variationally with respect to $E_0$
until the energy reached stability in a desired number of digits. 
The sequence of energies obtained for consecutive basis sets enables estimation 
of the basis truncation error.
The convergence for the $2^2P$ and $3^2S$ levels of B, and the $2^1S$ state of the B$^+$ ion 
is presented in Table I. 
The variational energies obtained from the largest expansion are lower than
the best  results previously reported in \cite{bubinB,BubinB+}.

\begin{widetext}

\begin{table}[!htb]
\renewcommand{\arraystretch}{1.3}
\caption{Convergence of the clamped nucleus energy $E_0$ (in a.u.) of the ground ($2^2P$) 
and the lowest excited ($3^2S$) states of boron atom. At the bottom, several 
results reported in literature are given.}
\label{T:Enr}
\begin{ruledtabular}
\begin{tabular}{rw{5.14}w{5.14}w{5.14}c}
Basis size  & \cent{2^2P}  & \cent{3^2S}  & \cent{2^1S({\rm B}^+)} & Ref. \\
\hline
1024               & -24.653\,755\,522     &  -24.471\,358\,195    & -24.348\,883\,829\,93  &  \\
2048               & -24.653\,844\,393     &  -24.471\,386\,316    & -24.348\,884\,352\,93  &  \\
4096               & -24.653\,864\,204     &  -24.471\,391\,933    & -24.348\,884\,458\,05  &  \\
8192               & -24.653\,867\,537     &  -24.471\,393\,366    &   \text{---}  &  \\
$\infty$           & -24.653\,868\,05(45)  &  -24.471\,393\,68(32) &  -24.348\,884\,479(14)   &  \\[1ex]
\hline                                                                   
5100               & -24.653\,866\,08(250) &  -24.471\,393\,06(50) &     & \cite{bubinB} \\
Full CI            & -24.653\,837\,33      &                       &     & \cite{almoradiazB} \\[1ex]
10000              &                       &                       & -24.348\,884\,446(35)   & \cite{BubinB+} \\
%\hline
% 8192 & -24.652\,500\,24(28)  &  -24.401\,941\,83(11) & $^{10} \rm B$ \\
% 8192 & -24.652\,623\,87(28)  &  -24.401\,817\,83(11) & $^{11} \rm B$ \\
\end{tabular}
\end{ruledtabular}
\end{table}

\end{widetext}

In a similar way, i.e. from the convergence with the growing basis set size,
the truncation errors of the expectation values of various operators were estimated.
Particular care was taken for the singular or nearly singular operators
$\left(p_a^4,\, \delta^3(r_{a}),\, \delta^3(r_{ab}),\, P(r_{ab}^{-3})\right)$,
which exhibit a slow numerical convergence of their mean values.
This undesirable effect is particularly pronounced for the ECG functions having improper 
short-distance behavior. The solution is to employ the regularized matrix elements 
following Drachman's recipes \cite{Sucher:79}. Previously \cite{PCK05}, 
this approach enabled the accuracy
of the expectation values to be increased by several orders of magnitude.
%Pertinent expressions for the relativistic operators have been presented in Appendix A. 

The expectation values of all the operators involved in the determination 
of the fine structure states energy are collected in Table~\ref{T:Hfs}. 
All the entries are accompanied by their estimated uncertainty. 
Table~\ref{T:Efs} contains the $\alpha$-expansion components 
and the final values of the measurable quantities: 
the $3^2S_{1/2} - 2^2P_{1/2}$ transition energy, 
the fine-structure splitting, and the ground state ionization potential 
for the most abundant $^{11}$B isotope  of the boron atom. In the table,
the theoretical predictions are compared with recent calculations 
\cite{ChenB,Derevianko,Froese_04} and
with the experimental values collected by Kramida and Ryabtsev \cite{KramidaBor}.
The agreement of the new values with the experimental results is apparent
and the remaining discrepancies are consistent with 
estimated uncertainties due to the approximate value of the Bethe logarithm 
and due to neglected higher-order $\mathcal{O}(\alpha^6)$ corrections.
In contrast, significant differences are observed with all the previous calculations. 
Although none of the cited theoretical values caries uncertainty,
it is clear that the number of digits quoted there is by far too high.
One may conclude, that the standard configuration interaction (CI), 
multiconfiguration Dirac-Fock (MCDF) or coupled clusters (CC) methods based 
on one-electron functions are not capable of supplying results with controlled precision.

The numerical results for the hyperfine splitting are presented in Tables~\ref{T:Hhfs}
and~\ref{T:Ehfs}. The former table collects the expectation values of individual
operators comprising the $H_\text{hfs}$ Hamiltonian and include
the Fermi contact, orbital term, spin-dipole term as well as
the term describing the interaction of the nuclear electric quadrupole moment
with the electric-field gradient produced by electrons. 
The head of this table presents a relation of the expectation values to the commonly 
used hyperfine parameters and the corresponding prefactors used in their evaluation.
We observe significant discrepancies for individual contributions in Table \ref{T:Hhfs}
with the results from the previous calculations by Chen \cite{ChenB}. 
This is particularly pronounced in the case of the Fermi-contact parameter $a_c$, 
for which this discrepancy is over 25\,\%.
Surprisingly, the differences between Chen's results and the previously established 
experimental $A$-hyperfine constants are much smaller (see Tab.~\ref{T:Ehfs}). Moreover, 
the difference between our result and the experimental values is consistent 
with the $O(Z\,\alpha)^2$ unknown relativistic correction, namely it is
about 50\% of $(Z\,\alpha)^2$ times the corresponding $A$ or $B$ coefficient.  
The final values for both the $2^2P$ levels and both $^{11}$B and $^{10}$B isotopes
are presented in Table \ref{T:Ehfs}.

\begin{widetext}

\begin{table}[!hbt]
\renewcommand{\arraystretch}{1.3}
\caption{Expectation values of various spinless and fine-structure operators for
$2^2P$ and $3^2S$ states of boron atom. To obtain the mean values for $2^2P_J$, 
the value for $2^2P$ state has to be multiplied by the relevant $\{K_{1/2},K_{3/2}\}$ 
coefficient in curly bracket i.e. $\{1,-\frac{1}{2}\}$ following Eqs.~(\ref{E:Q})~--~(\ref{E:Qij}).}
\label{T:Hfs}
\begin{ruledtabular}
\begin{tabular}{cw{5.12}w{5.11}w{5.16}}
Operator & \cent{2^2P}  &\cent{3^2S} & \cent{2^1S({\rm B}^+)}\\
\hline
$H_0 $
& -24.653\,868\,1(5)  &  -24.471\,393\,7(3) & -24.348\,884\,479(14) \\
$\vec p_a \cdot \vec p_b$
& 0.271\,175(2)  &  0.607\,784(3) & 0.595\,137\,52(4) \\
$p_a^4$
& 5\,546.924(3)  &  5\,602.919(2) & 5598.710\,4(3) \\
$\delta(r_a)$
& 71.864\,97(3)  & 72.544\,82(2)  & 72.506\,327(3) \\
$\delta(r_{ab})$
& 3.538\,453\,2(14)  & 3.582\,111\,2(7)  & 3.577\,866\,33(12) \\
$p_a^i \bigl(\frac{\delta^{ij}}{r_{ab}}\!+\!\frac{r^i_{ab}\,r^j_{ab}}{r^{3}_{ab}}\bigr) p_b^j$
& 2.171\,606(3)  &  2.995\,875(3) & 3.025\,198\,47(14) \\
$\frac{\vec r_a}{r_a^3}\times\vec p_a\cdot\vec\sigma_a$
& -1.494\,336(5)  &  \text{---} & \text{---} \\
$\frac{\vec r_{ab}}{r_{ab}^3}\times\vec p_a\cdot\vec\sigma_a$
& -2.855\,953(3)  &  \text{---} & \text{---} \\
$\frac{\vec r_{ab}}{r_{ab}^3}\times\vec p_b\cdot\vec\sigma_a$
& 0.568\,459(4)  &  \text{---} & \text{---} \\
$P(r_{ab}^{-3})$
& -27.874\,22(6)  & -29.391\,67(4) & -29.459\,453\,0(3) \\
$\ln k_0$
& 6.195(5)^a  & 6.195(5)^a & 6.194\,4(9)^b \\
\end{tabular}
\begin{flushleft}
$^a$ estimated from $\ln k_0[2^1S({\rm B}^+)]$\\ 
$^b$ Ref. \cite{BubinB+} 
\end{flushleft}
\end{ruledtabular}
\end{table}

%\end{widetext}
%\begin{widetext}

\begin{table}[!thb]
\renewcommand{\arraystretch}{1.3}
\caption{Components of the $3^2S_{1/2} - 2^2P_{1/2}$ transition energy, the fine structure 
splitting, and the ionization potential (IP) for $^{11}$B atom and of the ground level energy
for $^{11}$B$^+$. For comparison, previous theoretical
predictions and the experimental results are given at the bottom. All entries in $\icm$.}
\label{T:Efs}
\begin{ruledtabular}
\begin{tabular}{cw{8.14}w{6.14}w{6.14}w{8.7}}
Component    & \cent{3^2S_{1/2} - 2^2P_{1/2}} & \cent{2^2P_{3/2} - 2^2P_{1/2}}  & \cent{\mathrm{IP}(2^2P_{1/2})} & 
\cent{2^1S_0({\rm B}^+)} \\
\hline
$m\,\alpha^2$          &   40\,048.50(5)  &              &   66\,936.15(5)    &  -5\,343\,962.445(3)    \\
$m\,\alpha^2 \,\eta$   &         1.686(0) &              &         0.207\,6(0)&           272.846(0)  \\
$m\,\alpha^4$          &        -12.424(3)& 15.287\,8(1) &      -10.131(2)    &         -1410.050(0) \\
$m\,\alpha^5$          &         1.66(24) &              &         1.59(20)   &           173.68(4)     \\
$m\,\alpha^6$          &         0.10(3)  & 0.000(2)     &         0.10(3)    &            10.851(0) \\
Total                  &   40\,039.52(24) & 15.288(2)    &   66\,927.91(21)   &   \\[1ex]
%\hline 
Theory, 2011 \cite{bubinB}$^a$        &   40\,049.887(200)  \\   
Theory, 2015 \cite{ChenB}         &   40\,008.67     & 15.523       &                    \\
Theory, 2012 \cite{Derevianko}    &   39\,892.82     & 19.75        &   66\,886.5  8     \\
Theory, 2004 \cite{Froese_04}     &   40\,005.27     & 15.39        &                      \\ [1ex]
Experiment \cite{KramidaBor}  &   40\,039.656(3) & 15.287(3)    &   66\,928.0  36(22)
\end{tabular}
\begin{flushleft}
$^a$ without relativistic and QED corrections 
\end{flushleft}
\end{ruledtabular}
\end{table}

\begin{table}[!htb]
\renewcommand{\arraystretch}{1.3}
\caption{Expectation values of hyperfine splitting operators for $2^2P$ state 
of boron atom $^{\infty}$B, in relation to standard hyperfine parameters.  
%The mean value $\frac{1}{J (J + 1)} \langle\vec J \ldots \rangle $ for $2^2P_J$, 
%includes the factor $\{K_{1/2},K_{3/2}\}$ following Eqs.~(\ref{E:Q})~--~(\ref{E:Qij}).
%All digits are significant or otherwise noted. 
}
\label{T:Hhfs}
\begin{ruledtabular}
\begin{tabular}{lw{3.12}w{5.11}w{3.10}w{3.8}}
Reference & \cent{ \vec \sigma_a \,\delta^3(r_a)\;(\equiv \frac{a_c}{4 \pi})} 
 &  \cent{ \frac{\vec r_a}{r_a^3}\times\vec p_a\;(\equiv -2\,a_l)} 
& \cent{\frac{\sigma_a^i}{r_a^3}\,
\Big(\delta^{ij}-3\frac{r_a^i\,r_a^j}{r_a^2}\Big)\;(\equiv 10\,a_{sd})} & 
\cent{\frac{1}{r_a^3}\,
\Big(\delta^{ij}-3\frac{r_a^i\,r_a^j}{r_a^2}\Big)\;(\equiv 5\,b_q)}\\ \hline
This work                        & 0.007\,536(2) & -1.560\,155(5)  & -1.677\,193(11) & -1.417\,48(12) \\[1ex]
Theory, 2015 \cite{ChenB}$^a$    & 0.010\,13      & -1.557\,6    & -1.684      & -1.400\,5 \\
Theory, 1993 \cite{JonssonB}$^a$ & 0.006\,828     & -1.561\,4    & -1.672      & -1.422
\end{tabular}
\begin{flushleft}
$^a$ values calculated for $^{11}$B 
\end{flushleft}
\end{ruledtabular}
\end{table}
  
%\end{widetext}
%\begin{widetext}

\begin{table}[!htb]
\renewcommand{\arraystretch}{1.3}
\caption{Hyperfine splitting parameters (in MHz) for the ground state $2^2P$ state; 
         relativistic and finite mass corrections are not included so the uncertainties are purely numerical; 
         magnetic moments are \cite{Stone:05} 
         $\mu(^{11}{\rm B}) = 2.688\,648\,9(10)\,\mu_\mathrm{N}$ and 
         $\mu(^{10}{\rm B}) = 1.800\,644\,78(6)\,\mu_\mathrm{N}$.}
\label{T:Ehfs}
\begin{ruledtabular}
\begin{tabular}{lw{5.8}w{5.8}w{5.8}w{5.7}w{5.7}w{5.7}}
Reference & 
\cent{A_{1/2}(^{11}{\rm B})}  & \cent{A_{3/2}(^{11}{\rm B})} &\cent{B_{3/2}(^{11}{\rm B})} &
\cent{A_{1/2}(^{10}{\rm B})}  & \cent{A_{3/2}(^{10}{\rm B})} &\cent{B_{3/2}(^{10}{\rm B})} \\
\hline
This work                  & 365.710\,1(18) & 73.395\,3(15) & 2.704\,0(4)  & 122.462\,4(6) &  24.577\,8(6)  & 5.635\,1(8) \\[1ex]
Theory, 2015 \cite{ChenB}        & 365.91         & 73.41         & 2.675        & 122.21        &  24.92         & 5.575       \\
Theory, 2012 \cite{Derevianko}   & 373.3          & 72.7          &              &               &                &             \\
Theory, 1996 \cite{Jonsson_96}   & 366.1          & 73.24         &              &               &                &             \\
Experiment, 1960 \cite{Lew:60}   & 366.076\,5(15) &               &              & 122.585\,1(9) &                &    \\         
Experiment, 1972 \cite{Harvey:72}&                & 73.349\,6(4)  & 2.692\,7(10) &               &                &             
\end{tabular}
\end{ruledtabular}
\end{table}

\end{widetext}

%\onecolumngrid\newpage\twocolumngrid

\section{Conclusions}
We calculated the energy levels, isotope shifts, and fine and hyperfine structure in the atomic boron 
with numerical precision of a few parts per million. We demonstrated that the majority of the previous 
calculations were not as accurate as claimed---instead of five to six digits only the first two were significant.
This is particularly apparent for the fine structure splitting and for the Fermi contact
interaction, the last one being exceptionally small for the $2P$ ground state. 
The smallness of the Fermi contact term makes the hyperfine splitting insensitive to the
not-well-known nuclear finite size effects, thus the comparison with experimental hfs 
will be a good test of the atomic computational methods.
Moreover, the precise value for the mass polarization correction (see Tab.~\ref{T:Hfs}) permits 
the accurate determination of the isotope shift in the $2P-3S$ transition,
which paves the way for determination of the nuclear charge radius
of the proton halo in the $^8$B nucleus,
as we have already demonstrated for the beryllium atom \cite{Puchalski:14a}.

\section*{Acknowledgments}
This research was supported by National Science Center (Poland) Grant No. 2014/15/ B/ST4/05022 
as well as by a computing grant from Pozna\'n Supercomputing and Networking Center and by PL-Grid Infrastructure.

%\bibliography{Be}

\end{document}